\documentclass[amsmath, amssymb, prl, twocolumn]{revtex4-1}
\usepackage{graphicx}

\begin{document}
\title{Machine Learning 1- and 2-electron reduced density matrices of polymeric molecules}
\author{David Pekker}
\affiliation{Creyon Bio, 3210 Merryfield Row San Diego, CA 92121}

\author{Chungwen Liang}
\affiliation{Creyon Bio, 3210 Merryfield Row San Diego, CA 92121}

\author{Sankha Pattanayak}
\affiliation{Creyon Bio, 3210 Merryfield Row San Diego, CA 92121}

\author{Swagatam Mukhopadhyay}
\affiliation{Creyon Bio, 3210 Merryfield Row San Diego, CA 92121}

\begin{abstract}
Encoding the electronic structure of molecules using 2-electron reduced density matrices (2RDMs) as opposed to many-body wave functions has been a decades-long quest as the 2RDM contains sufficient information to compute the exact molecular energy but requires only polynomial storage. We focus on linear polymers with varying conformations and numbers of monomers and show that we can use machine learning to predict both the 1-electron and the 2-electron reduced density matrices. Moreover, by applying the Hamiltonian operator to the predicted reduced density matrices we show that we can recover the molecular energy. Thus, we demonstrate the feasibility of a machine learning approach to predicting electronic structure that is generalizable both to new conformations as well as new molecules. At the same time our work circumvents the N-representability problem that has stymied the adaption of 2RDM methods, by directly machine-learning valid Reduced Density Matrices. 
\end{abstract}
\maketitle

The methods of quantum chemistry, like density functional theory (DFT) and coupled-clusters methods, are key to \emph{ab initio} understanding of molecular properties. However, these methods are slow. Traditional implementations of the less accurate DFT methods scale as $n^3$, where $n$ is the number of electrons. The more accurate CCSD (coupled-clusters singles-doubles) method scales as $n^6$, and CCSD(T), which also includes triples scale as $n^8$. It is important to recognize that while CCSD(T) is currently considered to be the gold standard of quantum chemistry it still involves major approximations which preclude it from describing strongly correlated systems like Fractional Quantum Hall Effect~\cite{Laughlin1983}. Nevertheless, making use of the fact that quantum correlations are essentially local, i.e. the quantum nearsightedness principle~\cite{Kohn1995,Kohn1996}, the latest generation of quantum chemistry methods are approaching linear scaling with the number of electrons~\cite{Yang1991, Li1993, Goedecker1994, Hernandez1996, Onetep2020, BigDFT2020, Hampel1996, Schutz2000, Riplinger2013, Orca2020}. However, these tools are still too slow to compute quantum mechanical observables on truly large molecules like proteins and nucleic acids. 

The traditional method to overcome the limitations of \emph{ab initio} quantum chemistry has been to use semi-empirical approaches like the classical force-fields that are commonly used in molecular dynamics~\cite{Adcock2006}. Recently, there has been considerable work on using machine learning (ML) to develop much more accurate models of molecules~\cite{Unke2021}. The main goal of this line of ML research has been on building models that predict properties like molecular energies and forces on the nuclei significantly faster than the high-level methods of quantum chemistry, while simultaneously being significantly more accurate compared to the semi-empirical models used in classical force-fields.

Going further, several recent papers argued that machine learning can be applied directly to learning the characteristics of the electronic structure of molecules. These efforts have looked at predicting the one-electron Hamiltonian~\cite{Hegde2017, Schutt2019} and the 3D electron density~\cite{Unke2021c, Rackers2022, lee_rackers_bricker_2022}.

In this paper we demonstrate that machine learning can be used to directly predict the electronic structure of molecules. Specifically, we show that all desired 1- and 2-electron correlations can be predicted at any level of theory from Hartree-Fock and DFT to CCSD and beyond. One of our key innovation, compared to previous work, is representing electronic structure using reduced density matrices (RDMs) in the context of machine learning. 

First, we argue that representing electronic structure using reduced density matrices is advantageous to alternative representations like wave functions. Specifically, the sequence of $n$-electron reduced density matrices ($n$-RDMs) forms a hierarchy of complexity that encodes correlations between more and more electrons as $n$ increase. For example the 2RDM, which is obtained by tracing the full electronic reduced density matrix over all electron coordinates but 2, encodes correlations between 2 electrons. Crucially, physical observables tend to involve only few electron correlations and hence we can almost always truncate the sequence at $n=1$ or $n=2$. The 1-electron reduced density matrix (1RDM) is sufficient to compute observables like the electron density, dipole moment, and Hartree-Fock or DFT energy functional. The 2RDM is sufficient to compute the exact molecular energy, at any level of theory, as the Hamiltonian of quantum chemistry involves at most pairwise interactions between electrons~\cite{Lowdin1955, Mayer1955}. Further, as a consequence of quantum nearsightedness~\cite{Kohn1995, Kohn1996}, RDMs tend to have local support. This notion of locality has two critical consequently: (1) the storage requirements for RDMs scale linearly with the number of atoms; (2) machine learned models of RDMs are transferable between different molecules as long as the training set contains the various chemical neighborhoods.

Second, we perform a set of proof-of-concept calculations to show that machine learning can indeed be used to predict 1- and 2-RDMs. For these calculations, we limit the scope of the problem to axial deformations of linear polymers: polyethylene, poly-amide, and polyproline-II. The purpose of this limitation is to avoid the additional complication of implementing equivarient rotations of atomic orbital basis between the lab frame and the local molecular frame, which has been investigated elsewhere in the literature~\cite{Unke2021b, Rackers2022, lee_rackers_bricker_2022}. We build datasets for conformations of polyethylene, poly-amide, and polyproline-II molecules of different numbers of monomers. We verify that the 1- and 2-RDMs of these molecules indeed show quantum nearsightedness. Finally, we proceed to show that fixed-length descriptors, similar to Ref.~\cite{Pronobis2018}, can be used to machine learn both the 1- and the 2-electron reduced density matrices (1RDMs and 2RDMs). Specifically, we demonstrate that the resulting models can be used to construct 1RDMs and 2RDMs for molecular conformations that were not part of the training set. Moreover, we show that the 1RDMs and 2RDMs learned on shorter polymers can be used to construct 1RDMs and 2RDMs for longer polymers. Finally, we use the example of molecular energy to show that the RDMs constructed by our model can be used to accurately compute molecular properties. 

\section{Results}
\subsection{Representing electronic structure using reduced density matrices}
Typically, electronic structure in chemistry is represented in terms of a wave function that is made up of one or more Slater determinants. This type of representation suffers from two problems. First, wave functions are not local and hence there is no obvious way to ``glue" together wave functions of fragments in order to construct the wave function of a bigger molecule. Second, if one desires to go beyond the mean field level, the amount of information required to store the wave function grows very rapidly with the system size.  In the following we will argue that RDMs address both of these problems and therefore are the right representation for machine learning electronic structure.

We will exclusively work with closed shell molecules and therefore we will use spin-traced RDMs throughout. The hierarchy of the RDMs begins with the spin-traced 1RDM
\begin{align}
    \rho_{ij}^{(1)} &= \sum_{\alpha \in \{\uparrow,\downarrow\}} \langle c^\dagger_{\alpha j} c_{\alpha i} \rangle,
\end{align}
where $i$ and $j$ are atomic orbital indices, $\alpha$ is a spin index, $c^\dagger_{\alpha j}$ is the electron creation operator and $c_{\alpha i}$ the electron destruction operator. The next member of the hierarchy is the connected component of the spin-traced 2RDM
\begin{align}
     \tilde{\rho}_{ijkl}^{(2)} &= \sum_{\alpha, \beta, \gamma, \delta \in \{\uparrow,\downarrow\}} \langle c^\dagger_{\alpha j} c_{\beta i} c^\dagger_{\gamma l} c_{\delta k} \rangle - \rho_{ij}^{(1)} \rho_{kl}^{(1)}+\frac{1}{2} \rho_{il}^{(1)}  \rho_{kj}^{(1)}, \label{eq:2RDMdef}
\end{align}
where the factor of $1/2$ in the last term is a consequence of tracing over spins. We have removed the disconnected component of the 2RDM as it describes two  independent electrons and therefore does not have local support. The disconnected component of the 2RDM can be reconstructed, if needed, using the 1RDM which does have local support. Higher members of the RDM hierarchy can be constructed along similar lines if they are needed. 

Next, we address the problem of composing electronic structure of a bigger molecule from electronic structure of fragments. The principle of quantum nearsightedness, as described by Kohn~\cite{Kohn1995, Kohn1996}, states that electron-electron correlations in molecules tend to be short-ranged. A direct consequence of quantum nearsightedness is that 1RDMs have vanishingly small matrix elements corresponding to spatially well-separated atomic orbitals. This feature of 1RDMs underlies the linear scaling DFT methods implemented in packages like BigDFT~\cite{BigDFT2020} and ONETEP~\cite{Onetep2020}. Quantum nearsightedness implies that not just 1RDMs, but connected-components of the $n$-RDMs have local support. The essence of the present paper is the notion that machine learning can be used to predict the RDMs of fragments which can be glued together into the RDM of the bigger molecule much in the same way as Khon-Sham theory is used to predict 1RDMs of fragments that are glued together in divide-and-conquer linear-scaling DFT\cite{Shimojo2008}.

We comment that the principle of quantum nearsightedness is closely related to the area-law scaling of quantum entanglement of gapped quantum systems~\cite{Hastings2007, Eisert2010}. As most macro-molecules are indeed insulating and have a gapped spectrum, it is reasonable to expect that all electron-electron correlations in these molecules will decay exponentially with distance~\footnote{There are well known exceptions to quantum nearsightedness, like magnetic and superconducting systems that have simple long-range correlations, conducting molecules and metals that have power-law decaying correlations. Notions described in this paper would need to be suitably adapted for these cases, e.g. by taking special care of the long-range order or working in momentum space.}. The notion that the 1RDMs tend to decay exponentially with distance has been established in Refs.~\cite{Kohn1973, Taraskin2002}. We verify the quantum nearsightedness of 1- and 2-RDMs of polyethylene, as well as 1RDMs of poly-amide, and polyproline-II in a later subsection. 

Since at least 1955~\cite{Lowdin1955, Mayer1955}, there has been an ongoing interest in representing electronic structure using the 2-electron reduced density matrix (2RDM). As the Hamiltonian of quantum chemistry (at the level of the Born approximation) involves only pairwise interactions between electrons, both the many-body wave function and the 2RDM contain sufficient information to obtain the exact electronic energy. However, the storage requirements are vastly different: the many-electron wave function requires exponential amount of storage while the 2RDM only needs a polynomial amount of storage. While the full 2RDM requires $m^4$ storage, where $m$ is the basis size, the notion of quantum nearsightedness states that for large molecule the storage requirement actually scales linearly with $m$. Despite the vast advantages in storage, 2RDMs are not routinely used for quantum chemistry calculations due to the so-called $n$-representability problem: given a candidate matrix there is no computationally efficient method for checking whether it is a valid 2RDM, i.e. a 2RDM derived from an $n$-electron wave function. We point out that approximate validity checking does show promise~\cite{Mazziotti2006}. Machine learning offers an alternative route to bypass the $n$-representability problem whereby the features of valid 2RDMs are learned from the training data generated by conventional, but costly, wave function approaches like CCSD.

\subsection{Polymer datasets for machine learning reduced density matrices}
In order to test machine learning of electronic structure, we have chosen three polymers: polyethylene, poly-amide, and polyproline. We wanted to avoid equivarient transformations associated with bends and twists of molecules--- all three polymers were chosen because they can be well approximated as quasi-1D molecules, aligned along the $z$-axis in our case. Polyethylene was chosen as the simplest example of an organic polymer. Poly-amide was chosen as one of the simplest biologically relevant polymers. The poly-amide polymer selected here is more complex than the poly-amide backbone of proteins or peptides. We used a simplified version of the peptide nucleic acid (PNA) antisense oligonucleotide backbone~\cite{Nielsen1999} by omitting the nucleobases. Finally, polyproline was chosen both because of its biological relevance in protein structures and also to show that our methods of machine learning electronic structure are applicable beyond the very simplest polymers. Polyproline-II is the dominant conformation in fibrillar proteins (e.g., collagen), often found in unfolded proteins and play critical role in signal transduction~\cite{Adzhubei2013}. Polyproline-II has a helical structure that makes one $360^{\circ}$ turn for every three monomers. This commensurate twist angle allows us to avoid the problems of equivarient transformations by introducing three different monomer, each rotated by  $120^{\circ}$ relative to the previous monomer.

\begin{figure*}
    \centering
    \includegraphics[width=\textwidth]{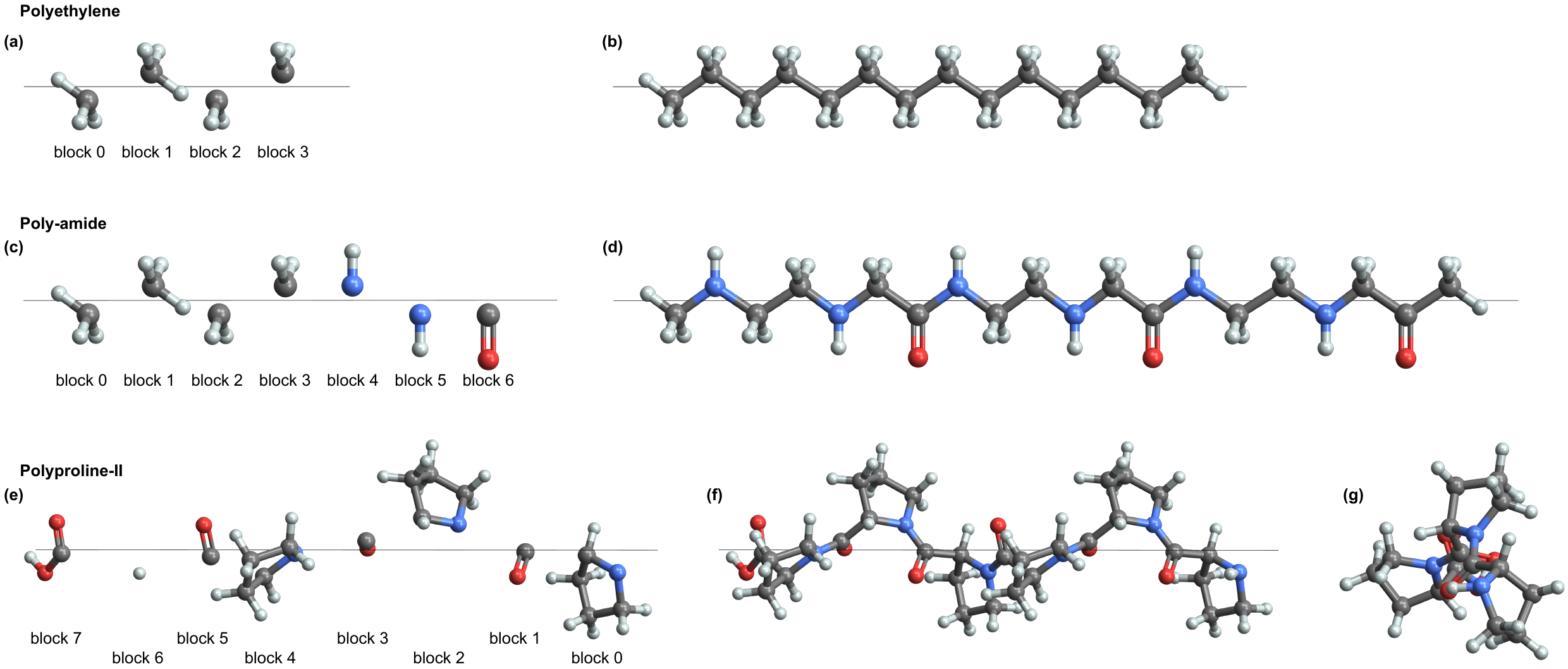}
    \caption{Polymer building blocks for (a) polyethylene, (c) poly-amide, and (e) polyproline-II.
    For each polymer, the first two building blocks correspond to the end-caps and the remaining building blocks are used to make up the interior part of the polymeric chain. The $z$ axis of the polymer is depicted as the thin black line. The polymers corresponding to the building blocks are depicted in (b),(d), and (f). (g) An end-on view of polyproline-II, showing the $120^{\circ}$ twist for every monomer.}
    \label{fig:building_blocks}
\end{figure*}

\begin{figure}
    \centering
    \includegraphics[width=\columnwidth]{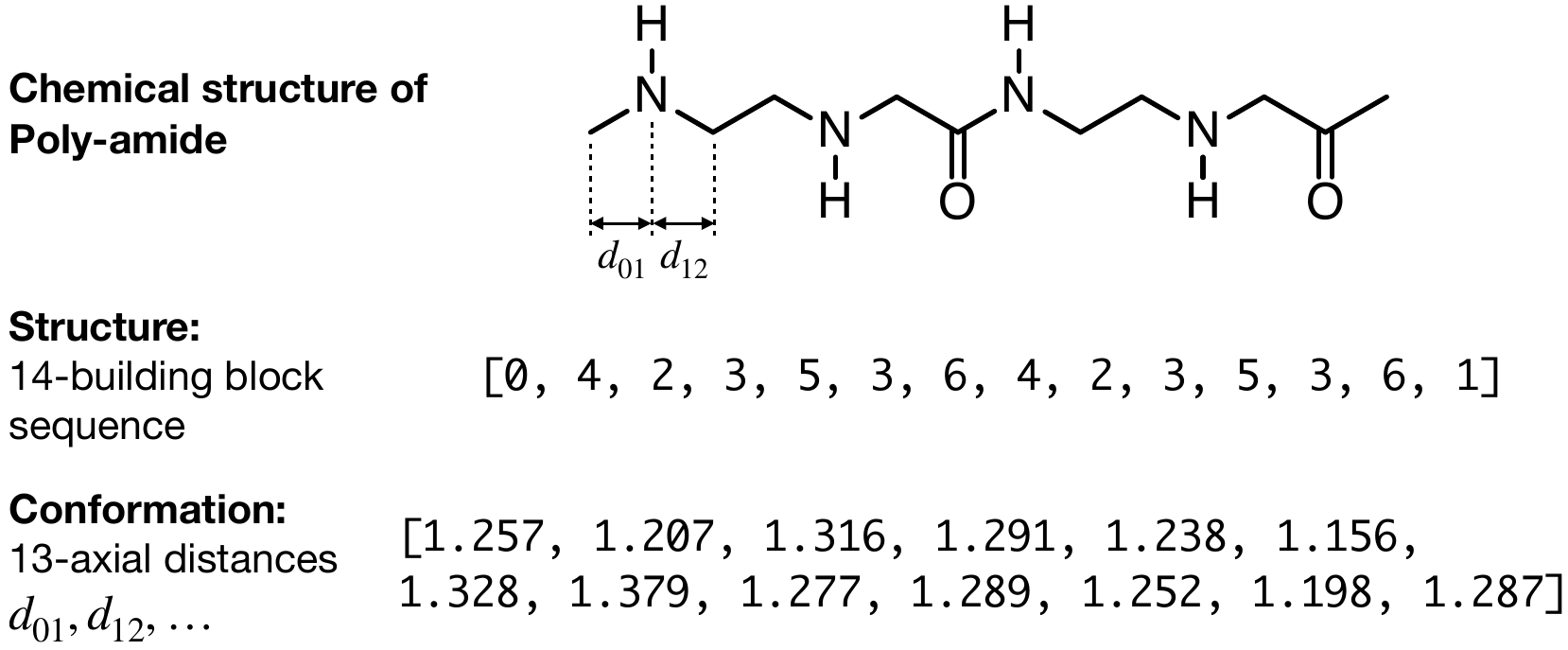}
    \caption{(a) Chemical structure of the poly-amide polymer. The monomer here is composed of two repeating subunits. (b) The structure is specified by a 14-building block sequence, using the building blocks depicted in Fig.~\ref{fig:building_blocks}c. (c) A particular conformation is specified using the 13 axial distances between the 14-building blocks that make up the polymer. The axial distance between the first pair of building blocks is denoted $d_{01}$,  between the second pair $d_{12}$, etc. }
    \label{fig:poly_amide}
\end{figure}

In order to specify the structure and conformation of a particular class of polymers, we begin by decomposing the polymer into a collection of building blocks as depicted in Fig.~\ref{fig:building_blocks}. The set of building blocks consists of a pair of end-caps as well as a number of building blocks that make up the bulk of the polymer. To specify the chemical structure of a polymer of a particular length, we specify the sequence of building blocks that make up that polymer. Finally to specify the conformation of the polymer, we specify the axial distances between the building blocks that make up the polymer (see Fig.~\ref{fig:poly_amide} for an example of specifying poly-amide 2). In addition to using building blocks to specifying the structure of a polymer, we will also use them to specify blocks of RDMs and to measure spatial separations between atomic orbitals.

\begin{table*}[]
    \centering
    \begin{tabular}{|l|l|l|c|c|c|c|c|c|c|}
        \hline
        Dataset name          & polymer & formula  & training & testing & RMSE $\rho^{(1)}$ & RMSE $\rho^{(2)}$ & RMSE $\text{Tr} \rho^{(1)}$ & RMSE $\text{Tr} \rho^{(2)}$ & RMSE Energy  \\
        \hline
        polyethylene          & PE-7    & $C_{14} H_{30}$               & 900 & 100 & $2.34*10^{-5}$ & --- & 0.00096 & --- & 0.22 kcal/mol\\
        DFT                   & PE-8    & $C_{16} H_{34}$               & --- & 50  & $2.13*10^{-5}$ & --- & 0.0013  & --- & 0.32 kcal/mol\\
                              & PE-9    & $C_{18} H_{38}$               & --- & 50  & $3.79*10^{-5}$ & --- & 0.0014  & --- & 0.37 kcal/mol\\
        \hline
        poly-amide            & PA-3    &  $C_{14} N_6 O_3 H_{30}$      & 900 & 100 & $2.32*10^{-5}$ & --- & 0.0019  & --- & 0.47 kcal/mol\\
        DFT                   & PA-4    &  $C_{18} N_8 O_4 H_{38}$      & --- & 50  & $4.27*10^{-5}$ & --- & 0.0043  & --- & 0.76 kcal/mol\\
                              & PA-5    &  $C_{22} N_{10} O_5 H_{46}$   & --- & 50  & $6.70*10^{-5}$ & --- & 0.0074  & --- & 0.78 kcal/mol\\
        \hline
        polyproline-II        & PPII-9  & $C_{45} N_{9} O_{10} H_{65}$  & 900 & 100 & $5.52*10^{-6}$ & --- & 0.00086 & --- & 0.17 kcal/mol\\
        DFT                   & PPII-12 & $C_{60} N_{12} O_{13} H_{86}$ & --- & 50  & $1.16*10^{-5}$ & --- & 0.0011  & --- & 0.19 kcal/mol\\        
        \hline
        polyethylene          & PE-5    & $C_{10} H_{22}$               & 900 & 100 & $3.77*10^{-7}$ & $4.85*10^{-7}$ & 0.00068 & 0.11 & 0.39 kcal/mol\\
        CCSD                  & PE-6    & $C_{14} H_{30}$               & --- & 50  & $1.67*10^{-6}$ & $6.24*10^{-6}$ & 0.0020  & 0.40 & 2.09 kcal/mol\\
        \hline
    \end{tabular}
    \caption{Datasets for machine learning of 1- and 2-RDMs. 
    The last five columns of the table summarize the deviation of machine learning predictions of various values away from quantum chemistry calculations performed on RDMs truncated at $r_q$. RMSE $\rho^{(1)}$ and RMSE $\rho^{(2)}$ measure the root mean square error (RMSE) over all of the predicted values of the reduced density matrix elements. RMSE $\text{Tr} \rho^{(1)}$ and RMSE $\text{Tr} \rho^{(2)}$ measure the RMSE of $\sum_{ij} \rho_{ij}^{(1)} S_{ij}-n$ and $\sum_{ijkl} \rho_{ijkl}^{(2)} S_{ij}S_{kl}-n(n-1)$, where $S_{ij}$ is the matrix of overlap integrals between atomic orbitals and $n$ is the number of electrons. RMSE Energy measures the RMSE between the energy computed with the predicted RDMs and the quantum chemistry computed RDMs.
    }
    \label{tab:datasets}
\end{table*}

We have generated a total of 4 datasets, that are summarized in Table~\ref{tab:datasets}. For each class of polymers, we picked a fixed length to generate the training data (e.g. for polyethylene we picked PE-7, i.e. $C_{14} H_{30}$). Using this fixed length we generated 900 training conformations and an additional 100 testing conformations. In order to test the transferability of our machine-learned models, for each class of polymer we also picked one or two longer fixed lengths (e.g. for polyethylene we picked PE-8, i.e. $C_{16} H_{34}$ and PE-9 , i.e. $C_{18} H_{38}$) and generated an additional 50 conformations of testing data. 

\subsection{Quantum nearsightedness of 1- and 2-electron reduced density matrices}
\label{sec:qns}
We will now verify that our datasets have the quantum nearsightedness property. We begin by looking at the structure of the 1RDMs, $\rho_{ij}^{(1)}$, and show that (1) the matrix elements decrease rapidly as a function of the distance between $i$ and $j$ and that (2) local conformation changes effect $\rho_{ij}^{(1)}$ only locally. Next, we verify that matrix elements of both 1RDMs and 2RDMs fall off exponentially with distance. Finally, we introduce a cutoff range on the RDM matrix elements and show that for range $\gtrsim 4$ the expectation value of the energy is only weakly effected by the cutoff. 

\begin{figure*}
    \centering
    \includegraphics{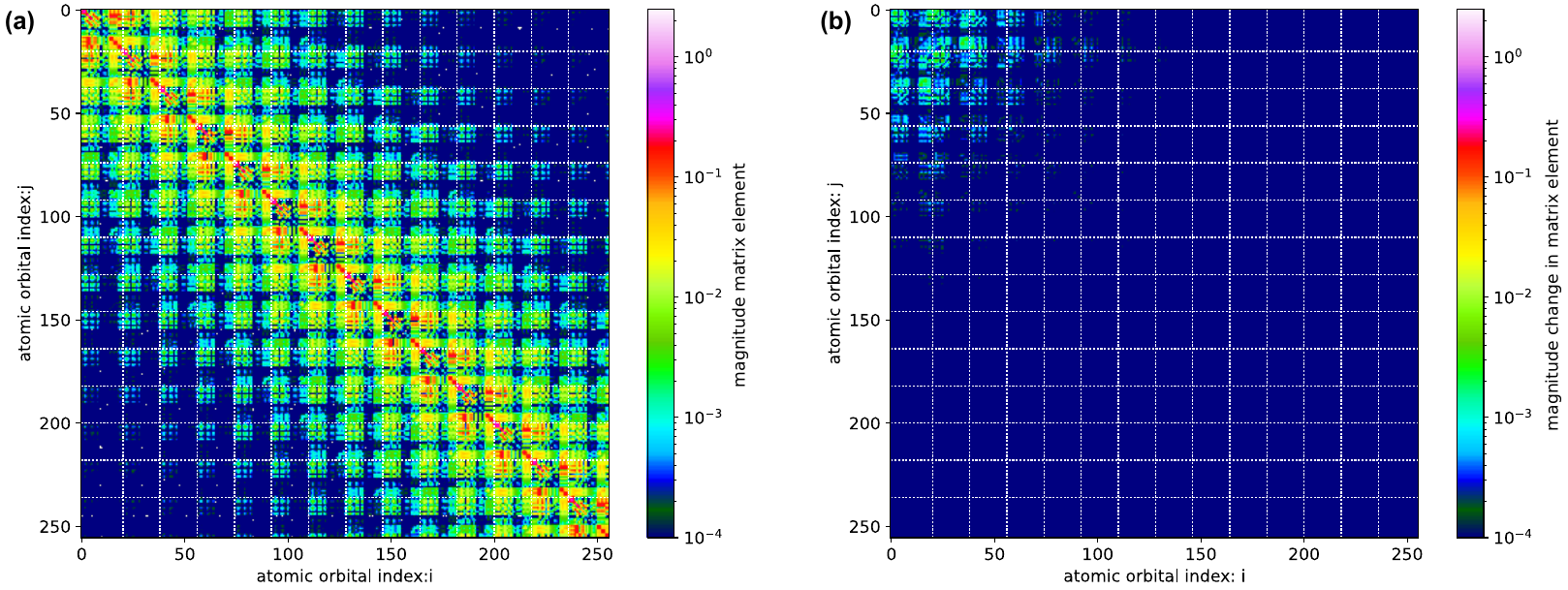}
    \caption{One electron reduced density matrices: (a) Decay in the matrix elements of the 1RDM, $\rho_{ij}^{(1)}$, as a function of separation between atomic orbitals $i$ and $j$. Depicted is the 1RDM of Polyethylene-7 obtained using DFT. Color indicates the magnitude of the matrix elements on a logarithmic scale. White dotted lines indicate the boundaries between building blocks. (b) Effect of a local conformation change on the 1RDM of Polyethylene-7. Depicted is the difference between the 1RDM of PE-7 with all carbon-carbon bonds of identical length and the 1RDM of PE-7 with the first carbon-carbon bond stretched (along the axial direction of the polymer) by $5\%$.  }
    \label{fig:1RDM_color}
\end{figure*}

\begin{figure*}
    \centering
    \includegraphics[width=0.9\textwidth]{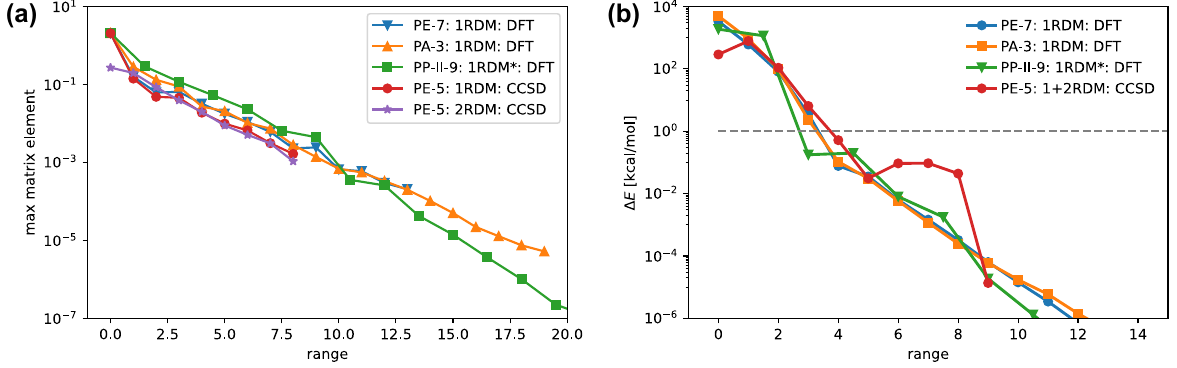}
    \caption{Quantum nearsightedness: (a) Maximum (by magnitude) RDM matrix element as a function of range computed using the building-block distance for a single molecule from each of our four datasets (as labeled). For PE-7, PA-3, and PP-II-9 the 1RDM was obtained using DFT; for PE-5 the 1RDM and the connected component of the 2RDM were obtained using CCSD. (b) Energy expectation value obtained from the truncated RDMs as a function of the truncation range. DFT energy functional was applied to the truncated DFT 1RDM for PE-7, PA-3 and PP-II-9. The full quantum chemistry energy functional was applied to the truncated CCSD 1RDM and 2RDM for PE-5. The dotted line indicates chemical accuracy at $1~\text{kcal}/\text{mol}$. To facilitate comparison between the different molecules, for PP-II-9 the range was stretched by a factor of 1.5 as half of its building blocks (blocks 0, 2, and 4) contribute two atoms to the backbone of the polymer while all other building blocks contribute only 1 atom to the backbone. }
    \label{fig:1RDM_decay}
\end{figure*}

In Fig.~\ref{fig:1RDM_color}a we plot the 1RDM for PE-7 in the atomic orbital basis. In constructing this plot, we have organized the basis by building block and then by atom, such that atomic orbitals on end of the polymer are in the upper left of the 1RDM and those on the other side on the bottom right. This ordering of atomic orbitals gives the 1RDM the characteristic block-like structure. We observe that the largest matrix elements are in blocks along the main diagonal -- these are the matrix elements that belong to the same atom. As we move away from the main diagonal we see rapid decay as we first encounter matrix elements that belong to the same building block, then to neighboring building blocks, then to next-nearest neighboring building blocks, etc. This rapid decay away from the diagonal occurs for all three polymers that we consider (see Supplement, Fig.~S1). Next, we introduce a local deformation of the molecule by changing the distance between building block 0 and block 1 and look at the difference between the 1RDM of the deformed molecule and the original 1RDM (see Fig.~\ref{fig:1RDM_color}(b)). We observe that the deformation only effects matrix elements in the upper left portion of the 1RDM. Supplement Fig.~S2 shows that other local deformations also result in local changes of the 1RDM. This observation, in conjunction with quantum nearsightedness, implies that one should be able to create local descriptors to model the 1RDM. 

Next, we introduce the building-block distance metric, $g_{i,j}$, to measure distance between atomic orbitals $i$ and $j$
\begin{align} 
g_{i,j} = B \,\,  \forall i \in a, \forall j \in b  
\end{align} 
where $B$ is an integer distance counting the number of building blocks separating (along the linear axis of the polymer) the building block indexed by $a$ and $b$. We use $a, b, p, q$ as block indices throughout the paper. The distance between building blocks $a$ and $b$ is similarly defined as $g_{a,b}$, and is equal to $B$. 

Using the notion of building-block distance metric, we can quantitatively investigate quantum nearsightedness. Specifically, we expect matrix elements of the RDMs to decrease exponentially with distance. To verify that this is indeed the case, for each RDM we plot the largest, by magnitude, matrix element as a function of the distance, see Fig.~\ref{fig:1RDM_decay}a. We indeed observe exponential decay for all three polymers that we consider (polyethylene, poly-amide, and polyproline-II), both for 1RDMs obtained using DFT and 1- and 2RDMs obtained using CCSD. Surprisingly, if we use the number of bonds along the backbone as the metric, then all of the traces roughly collapse, suggesting that there is a single quantum nearsightedness lengthscale for this class of polymers. To further validate truncation of RDMs, we compute the energy of the test polymers as a function of the truncation range $r_q$, see Fig.~\ref{fig:1RDM_decay}b. We find that in all cases the truncation error shrinks rapidly, and falls below chemical accuracy for range $\gtrsim 4$.

\subsection{Proof-of-concept machine learning of reduced density matrices}

\begin{figure*}
    \centering
    \includegraphics{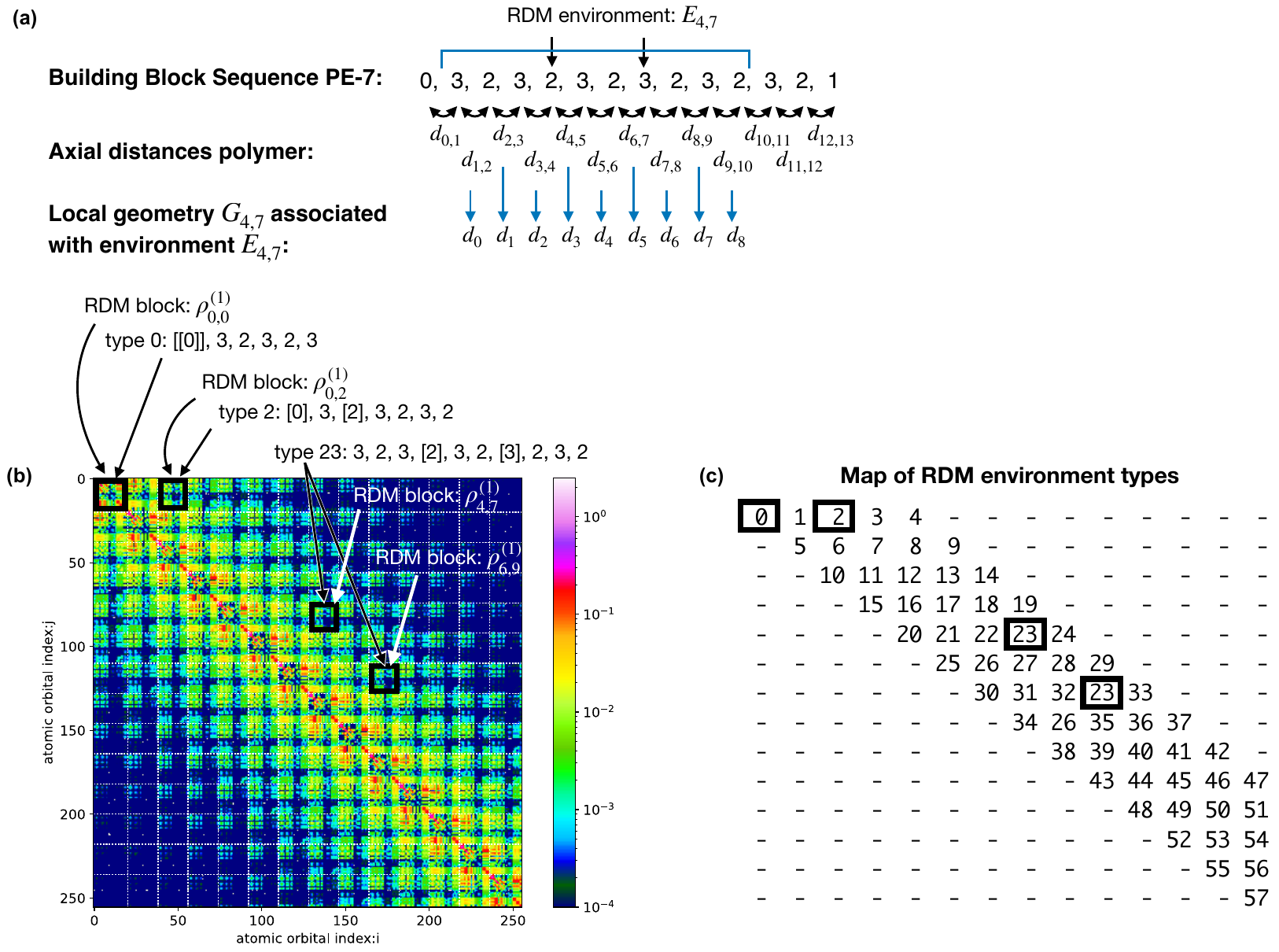}
    \caption{RDM blocks, RDM environments, and RDM environment types using polyethylene-7 as an example. (a) Sequence of building blocks that makes up polyethylene-7 (see Fig.~\ref{fig:building_blocks} for the definition of the building blocks) and axial distances that define its conformation. The location of the RDM environment $E_{4,7}$ and its local geometry $G_{4,7}$ are schematically depicted. (b) The 1RDM of polyethylene-7. Four RDM blocks belonging to three RDM environment types are highlighted. (c) Map of the 1RDM environment types. Values of RDM blocks in the lower part of the matrix are obtained by conjugating the corresponding blocks in the upper part. `-' indicates that no 1RDM environment type is assigned to the 1RDM block, either because it appears in the upper part of the matrix or because it is not required by quantum nearsightedness. }
    \label{fig:1RDM_environments_PE}
\end{figure*}

\begin{figure*}
    \centering
    \includegraphics[width=\textwidth]{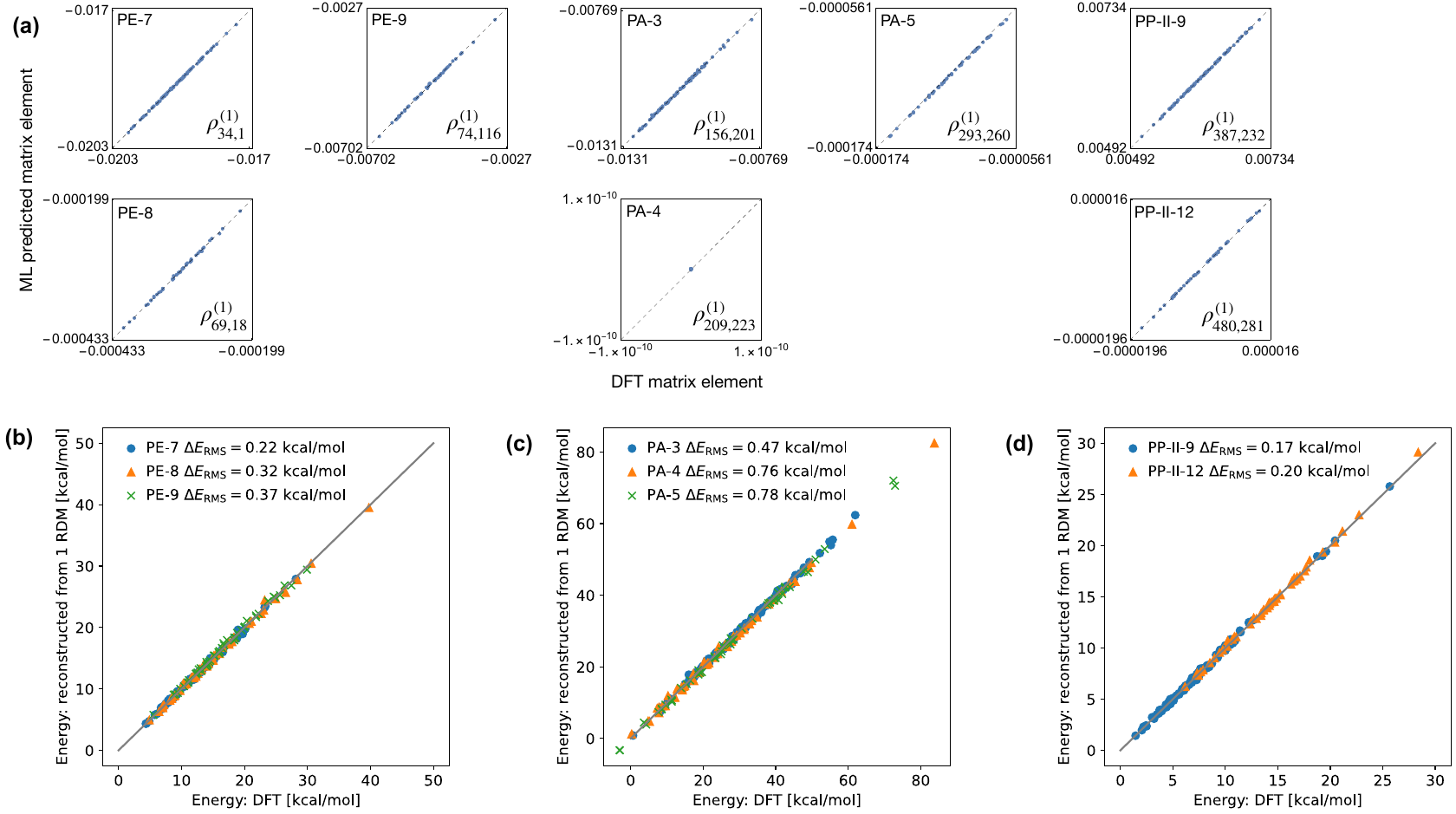}
    \caption{Machine learning 1RDMs produced by DFT calculations. (a) Comparison of RDM matrix elements predicted by machine learning to ones obtained by DFT. Each panel corresponds to a single, randomly chosen, matrix element sampled over all conformations in the corresponding testing dataset. The dashed line is a guide to the eye indicating perfect agreement between DFT and machine learning. Note, PA-4 matrix elements $\rho^{(1)}_{209,223}$ is zero to machine precision. (b,c,d) Comparison of the energy computed by DFT to the one obtained by evaluating the expectation value of the DFT energy functional on the predicted 1RDMs for polyethylene (a), poly-amide (b), and polyproline-II (c).}
    \label{fig:MLcomparisonDFT}
\end{figure*}

\begin{figure*}
    \centering
    \includegraphics[width=\textwidth]{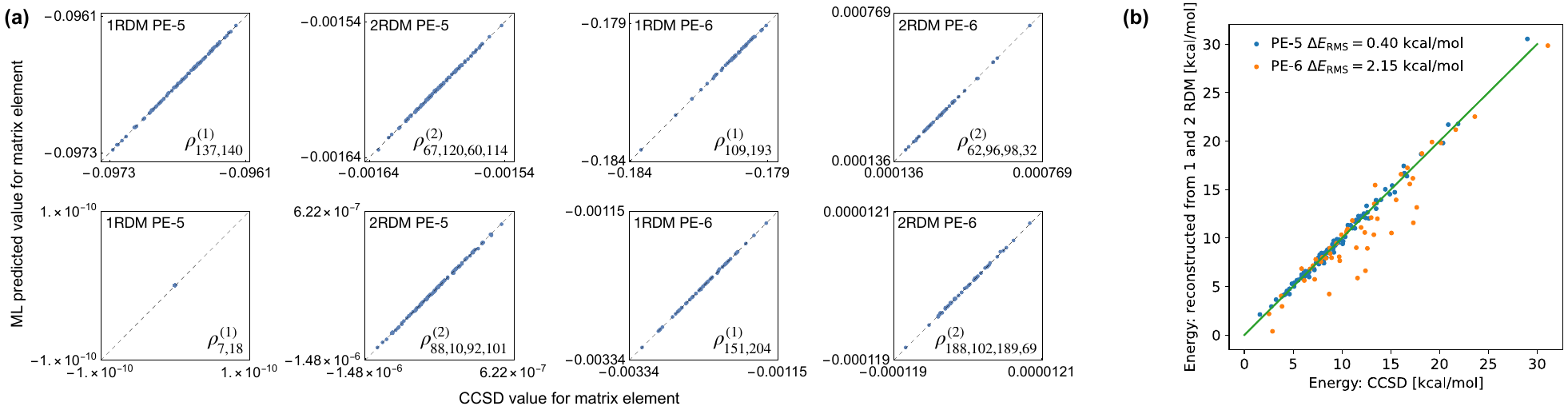}
    \caption{Machine learning 1RDMs and 2RDMs produced by CCSD calculations. (a) Comparison of RDM matrix elements predicted by machine learning to the ones obtained by CCSD. 
    Note, PE-5 matrix element $\rho^{(1)}_{7,18}$ is zero. (b) Comparison of the energy computed by CCSD to the one obtained by evaluating the expectation value of the quantum chemistry Hamiltonian on the predicted 1- and 2RDMs.}
    \label{fig:MLcomparisonCCSD}
\end{figure*}
Quantum nearsightedness implies: (property 1) both 1RDMs and the connected component of 2RDMs can be truncated at range $r_q \sim 4$ yielding a very good approximation for observables like energy; (property 2) local perturbations effect the connected components of RDMs only locally, within a range $r_c$. 

To construct our machine learning model, we need to define RDM blocks, RDM environments, and RDM environment types. An RDM block $\rho_{ab}^{(1)}$, is a submatrix of the RDM matrix, composed of all matrix elements $\rho_{ij}^{(1)}$ where atomic orbital $i$ is on building block $a$ and atomic orbital $j$ on building block $b$, Fig.~\ref{fig:1RDM_environments_PE}b. An RDM environment is an extension of the concept of chemical environments to the context of RDMs. An RDM environment $E_{ab}$ is associated with a particular RDM block $\rho_{ab}^{(1)}$ and consists of the set of building blocks that are in the neighborhood of building blocks $a$ and $b$, Fig.~\ref{fig:1RDM_environments_PE}a. Making use of property 1, we only consider short-ranged RDM environment $E_{ab}$, i.e. those with the building block distance $\text{g}_{a,b} \leq r_q$. For the DFT datasets, $E_{ab}$ consists of the list of building blocks $\{p\}$ such that the building block distance $\text{g}_{p,o}\leq r_c$ where $o$ is the mid-point between $a$ and $b$ and $r_c$ is the range introduced in property 2 (for the CCSD dataset we are forced to work with shorter polymers and hence use a slightly different definition of $E_{ab}$, see Methods). We associate an RDM environment type $E_{t}$, where $t$ is the type index, with RDM environments $\{E_{ab} \}$ that share: (1) building block types $a$ and $b$, (2) the distance between $a$ and $b$, and (3) the list of neighboring building block types, , Fig.~\ref{fig:1RDM_environments_PE}b. Crucially, distinct RDM blocks/RDM environments can be assigned the same RDM environment type (see Fig.~\ref{fig:1RDM_environments_PE}b\&c), which is precisely what allows us to apply models learned on short polymers to predict RDMs of longer polymers. 

We now define our machine learning model for the RDM blocks
\begin{flalign}
\rho_{ab}^{(1)}=
\left\{
\begin{array}{ll}
0 & \text{g}_{a,b} > r_q \\
{\cal P}^t (G_{ab}) & \textit{otherwise}
\end{array}
\right.\label{eq:model}
\end{flalign}
where $t$ is the RDM environment type index associated with the RDM block $a,b$; ${\cal P}^t$ is a polynomial fitting function associated with environment type $t$ that maps the local geometry $G_{ab}$ of the molecule to the submatrix of the RDM; and
\begin{align}
    G_{ab} = \{ d_{p,p+1}, d_{p+1,p+2}, ... \} = \{d_0, d_1, ...\}
\end{align} 
is a vector of the axial distances (in angstroms) between the building blocks of the environment $E_{ab}$. For brevity, we drop the two index notation for the axial distances, see Fig.~\ref{fig:1RDM_environments_PE}a. The elements of ${\cal P}^t$ are polynomials of 3rd-order over $G_{ab}$
\begin{align}
    {\cal P}^t_{ij} =& c^t_{ij,0} d_0 + c^t_{ij,1} d_1 + ... c^t_{ij,00} d_0^2 + c^t_{ij,01} d_0 d_1 + \nonumber \\
    &... c^t_{ij,000} d_0^2 + ...
\end{align}
where $c^t_{ij,.}$ are the fitting parameters that depend only on the RDM environment type (not the RDM environment) and the position of the element $ij$ in the RDM block. To determine the values of $c^t_{ij,.}$'s we use linear regression  (see Methods for details).

For the first test of our machine learning procedure, we use the 1RDMs from the three DFT datasets, see Table~\ref{tab:datasets}, to construct models for polyethylene (using PE-7 data), poly-amide (using PA-3 data), and polyproline-II (using PPII-9 data). Using these models, we predict 1RDMs for polymers of the same length as in the training data as well as for longer polymers. We perform three  types of characterization on the predicted 1RDMs. In the first characterization, we compare the predicted RDMs to ones obtained using DFT. In Fig.~\ref{fig:MLcomparisonDFT}a, we make the comparison for a group of randomly selected matrix elements (see supplement Figs.~S3-S5 for additional matrix elements and Figs.~S8-S9 for distribution of errors). In all cases we observe excellent agreement between the predicted and the DFT values of matrix elements. To quantify this error in the matrix elements, we compute the root mean square error (RMSE) between the predicted and the DFT matrix elements, see Table~\ref{tab:datasets} (see supplement for additional error quantification). We observe that RMSE is slightly higher for longer polymers than for shorter polymers, which is reasonable as the training set only consisted of the shorter polymers. For the second characterization, we verify a key property of the predicted 1RDMs: that the trace is equal to the number of electrons. In Table~\ref{tab:datasets} we list the RMSE of the trace of the 1RDMs and find errors on the order of 1/1000's of an electron, again with slightly larger errors for longer polymers. For the third characterization, we use the predicted 1RDM to compute the energy, see Fig.~\ref{fig:MLcomparisonDFT}b-d. The RMSE in energy are listed in Table~\ref{tab:datasets} and are well below chemical accuracy.

For the second test, we use 1RDM and 2RDM data obtained using CCSD to construct a model of polyethylene (using PE-5 data). In this case we work with shorter polymers as computer run time and storage become more expansive for larger polymers.  Each 2RDM in the learning set, following truncation at $r_c = 5$, still requires $\sim 2\text{Gb}$ and the the composition of the learning and training sets requires $\sim 3\text{Tb}$ of storage. We find that the agreement between predicted and computed matrix elements, see Fig.~\ref{fig:MLcomparisonCCSD}a, Table~\ref{tab:datasets}, and supplement Figs.~S6, S7, S9, is comparable for PE-5 but not quite as good for PE-6 as for the DFT datasets. The RMSE on the trace of the 2RDM, Table~\ref{tab:datasets}, is larger than RMSE on the trace of the 1RDM, which is partially because there are considerably more matrix elements to trace over in the 2RDM and partially because PE-5 is too short and hence the model does not transfer learning to PE-6 as well as for previous examples. Comparing energies computed CCSD to the ones obtained by applying the Hamiltonian of quantum chemistry to the predicted 1RDMs and 2RDMs, see Fig.~\ref{fig:MLcomparisonCCSD}b, we find good agreement for PE-5 but only reasonable agreement for PE-6. The RMSE in energy for PE-6, see Table~\ref{tab:datasets}, is slightly above chemical accuracy at $2.09\,\text{kcal/mol}$. In order to lower the RMSE in energy to this point, we had to devise a special procedure for extending the size of the local neighborhoods when defining the RDM environments (see methods), which was not needed for PE-7 training set that was used on the DFT data. Therefore, we ascribe the somewhat larger RMSE in energy of PE-6 to the imperfect transferability of the model from PE-5 due to PE-5 being too short (see Supplement). 

\section{Discussion}
We have demonstrated a pathway for machine learning electronic structure of polymers with accuracy that is essentially sufficient to reach chemical precision. Our key innovation is using reduced density matrices to encode the electronic structure. This innovation allows us to encode electronic structure at any level of theory from self-consistent field methods like Hartree-Fock and DFT to CCSD and beyond. We emphasize that unlike previous works~\cite{Hegde2017, Schutt2019,Unke2021c, Rackers2022, lee_rackers_bricker_2022}, our method allows us to machine learn much richer information about the electronic structure, specifically the electron-electron correlations. Learning these correlations allows us to evaluate expectation values of molecular operators with high precision. Indeed, for PE-5 the RMS error on the expectation value of the energy operator obtained from machine learned reduced density matrices as compared to CCSD density matrices is only $0.40~\text{kcal/mol}$, which is smaller than the RMS error comparing DFT energy to CCSD energy, of $0.81~\text{kcal/mol}$ (after removing the $675.83~\text{kcal/mol}$ systematic energy shift from the DFT energy). Moreover, the encoding is local in space which has two important benefits. First, our machine learned models are transferable between different molecules as long as the training set includes all reduced density matrix environments of the target molecule. Second, the size of our machine learned models scale at most linearly with the size of the molecule. For molecules with repeated patterns, like polymers, our models have a fixed size, independent of the polymer length.

The proof-of-concept calculations presented focused on linear polymers with both linear (polyethylene and poly-amide) as well as non-linear (polyproline-II) monomers. Using poly-amide, we have shown that complex polymers that are not composed of just one repeating subunit can be directly treated by our method. We expect that our method can treat linear co-polymers, block co-polymers, and random polymers by expanding the training set to span over the needed RDM environments. At the same time the method should be able to treat polymers with more complex monomers like linear (non-branched) sugars, nucleic acid backbones, and linear peptides with small side chains. Finally, we expect that our models are transferable to much longer polymers, composed of hundreds or thousands of monomers. However, we lack the quantum chemistry tools to validate such predictions. 

Next, we remark that our approach provides a method for performing quantum chemistry with 1- and 2RDMs, which has been discussed since at least 1955. 
Standard quantum chemical theories like Hartree-Fock, DFT, CCSD, etc make various approximations in how they treat electron correlations. Indeed, encoding all electron correlations at the level of the wave function requires exponentially large resources. On the other hand, because the Hamiltonian of quantum chemistry involves at most two-electron terms, the 1- and 2RDMs are capable of encoding all of the electron correlations that are needed to compute the electronic energy without making any approximations regarding electron correlations. Moreover, the storage requirements are essentially linear in the number of atoms as the RDMs are local due to quantum nearsightedness. However, the $n$-representability problem, the fact that there is no known computationally simple method to check if an ansatz 4-tensor is a valid 2RDM, has proved to be a significant impediment. Our approach allows one to bypass the $n$-representability problem using machine learning. Specifically, we use high-level theory to produce accurate 2RDMs for small polymers, then we machine learn what valid 2RDMs look like locally, and finally we take advantage of quantum nearsightedness to produce valid 2RDMs for large polymers by stitching together the local 2RDMs. 

In constructing 1- and 2-RDMs, we have made two assumptions. First, we can generate a sufficiently dense training set such that we can accurately learn local RDM environments. Crucially, for the case of 2RDMs, these local environments are compatible with n-representability.  Second, by stitching together locally valid 1- and 2-RDMs we can produce globally valid 1- and 2-RDMs that satisfy n-representability. For the proof-of-concept calculations that we describe in the present paper, we argue that we indeed find strong evidence to support both of these assumptions as the global 1- and 2-RDMs we generate using machine learning are indeed very close to the ones obtained by quantum chemistry calculations. Going forward, we expect that the size of the RDM environments, and hence the complexity of machine learning, could be reduced by explicitly considering electrostatic interactions mediated by classical electric fields. Speculating on a grander scale, we expect that if our assumptions hold true in general, then it should be possible to use  machine learning tools, like equivarient neural networks~\cite{Rackers2022,lee_rackers_bricker_2022,Unke2021b}, to extend our method from treating only axial deformations of linear polymers to arbitrary deformations of arbitrary molecules. We expect that it should be possible to construct a single unified machine learning model that, given a molecule and its conformation, predicts the molecule's electronic structure including electron-electron correlations, with an accuracy comparable to the state-of-the-art quantum chemistry tools, but does so in a fraction of the computational time-cost. 

\section{Methods}
\subsection{Generating quantum chemistry datasets}
For each molecule considered, all conformations were generated by randomly and independently picking axial distances between neighboring building blocks from a normal distribution centered on the nominal axial distance, with a standard deviation of $5\%$ of the nominal axial distance. For the three 1RDM datasets, we used DFT with b3lyp functional and 6-31G* basis. For the 2RDM dataset we used CCSD calculation with the 6-31G* basis. 

\subsection{1RDM neighborhoods for DFT data}

For DFT data we need to define 1RDM neighborhoods. Given a list of atomic orbitals $\{i\}$, we group the orbitals by building block $a$. We split the 1RDM, expressed in the atomic orbital basis, into RDM blocks $\rho^{(1)}_{ab}$ composed of matrix elements $\rho^{(1)}_{ij}$ such that $i \in a$ and $j \in b$. Applying the notion of quantum nearsightedness, we drop 1RDM blocks where the building-block distance between $a$ and $b$ is greater than $r_q=4$. Next, we assign the RDM block $\rho^{(1)}_{ab}$ a 1RDM environment type based on its neighborhood, i.e. the sequence of building blocks that are a distance $r_c$ from the the midpoint between blocks $a$ and $b$ along with the location of $a$ and $b$ in this sequence (where distances are measured using the building block distance metric). We use $r_c=5$ for polyethylene and polyproline-II and $r_c=6$ for poly-amide. We depict the assignment of 1RDM environment types for PE-7 in Fig.~\ref{fig:1RDM_environments_PE}. We can separate 1RDM environments into surface environments, like type 0 and type 2 environments, that bump into the end of the molecule on one side and bulk environments, like type 23 environment, that do not. A key property of our 1RDM environment assignment is that no new 1RDM environment types appear for longer polymers, instead previously defined bulk environments get used more and more times to describe the longer bulk of a longer polymer. 

\subsection{Machine learning DFT data}

Our training dataset provides us with a list of local deformations of 1RDM environments and the corresponding values of the 1RDM blocks. For each local local deformation we construct a non-linear descriptor as a polynomial function over the distances between the building blocks that belong to that environment. Specifically, let $\{d_0, d_1, ... d_{m-1}\}$ be the axial distances between the building-blocks of an $m$-building-block 1RDM environment. The descriptor we use has the form 
\begin{align}
    \mathcal{D}=& \{d_0, ..., d_{m-1}, \\
                & d_0^2, d_0 d_1, d_0 d_2, d_0 d_3, ..., d_{m-1}^2,\nonumber \\
                & d_0^3, ... , d_0 d_1 d_2, ..., d_{m-1}^3 \nonumber\}
\end{align}
where quadratic terms have a maximum separation of 3 building-blocks and cubic terms of 2 building-blocks. Next, we perform a separate linear regression for each element of the 1RDM block to determine its dependence on the descriptor $\rho_{ij}=c_{ij,0} + c_{ij,1} d_0 + c_{ij,2} d_1 + ...$ and write down the resulting coefficients $c_{ij,\mu}$.

To predict the value of a 1RDM given the geometry of the molecule, we first identify the 1RDM environments in the molecule. Then, for each environment we construct the polynomial descriptor. Using the descriptor and the stored coefficients we construct each 1RDM block. Finally we glue the 1RDM blocks into the density matrix.

\subsection{1RDM and 2RDM neighborhoods for CCSD data}

As explained in the main text, for the case of CCSD data we work with smaller molecules as the data requires significantly more resources to produce and store as compared to the DFT data. Specifically we use PE-5 for the training set and PE-5 and PE-6 for the testing sets. When we store the 1RDM and 2RDM data, we truncate the density matrix at $r_q=5$. Using shorter polymers for the training set constrains the size of neighborhoods in RDM environments. Indeed, these neighborhoods become too small to capture We attack the problem of this size constraint by being more flexible about how we pick neighborhoods so as to maximize their size. 

First, we define and learn both surface and bulk environments for the same RDM block. The surface environments tend to be larger than the bulk environments as they are allowed to include building-blocks at the end of the polymer. Hence when making predictions, we first attempt to assign an RDM block to a surface RDM environment and if that fails then we fall back to the bulk environment. This type of fallback routinely appears when transferring a model learned on shorter polymers to predicting the RDMs of longer polymers. 

Second, we maximize the size of the surface and bulk environments. Specifically, for surface environments in the training set we always pick the sequence of building blocks that begins on one side of the polymer and runs all the way to the other side but excludes the very last building block. For the bulk environments, we found the 'center' of the RDM block by averaging its indices (i.e. the midpoint between $a$ and $b$ for the 1RDM block $\rho^{(1)}_{ab}$ and the average of $a$, $b$, $c$, and $d$ for the 2RDM block $\rho^{(2)}_{abcd}$). We extended the local environment 3 building-blocks to the left and 4 building-blocks to the right if the mid-block was type 2; 4 building-blocks to the left and 3 building-blocks to the right if the mid-block was type 3. 

There were 10 2RDM blocks (out of 2712) in the PE-6 testing set that did not have matching 2RDM environment types from the learning set. In all 10 cases the range of the 2RDM blocks was maximal with $r_q=5$. These blocks were set to all zeros when predicting 2RDMs.

\subsection{Machine learning CCSD data}
The 1RDM and 2RDM environments for CCSD data were learned in almost the same way as the 1RDM environments for DFT data. The only difference was in the definition of the descriptor, we kept all quadratic terms up to a maximum separation of 5 building blocks and all cubic terms up to a separation of 1 building-block. 

\section{Acknowledgements}
The authors wish to thank Monica McArthur for helping us to get our code to run on the Google Cloud, Christopher E. Hart and the Creyon team for useful discussions and support, and Google Cloud for providing us with free compute time.

\bibliography{rdm}

\newpage
\onecolumngrid
\setcounter{figure}{0}
\let\oldthefigure\thefigure
\renewcommand{\thefigure}{S\oldthefigure}
\section{On-line supplement}
In this supplement we provide additional numerical data.

\subsection{Quantum nearsightedness in 1RDMs}

\begin{figure*}[h]
    \centering
    \includegraphics{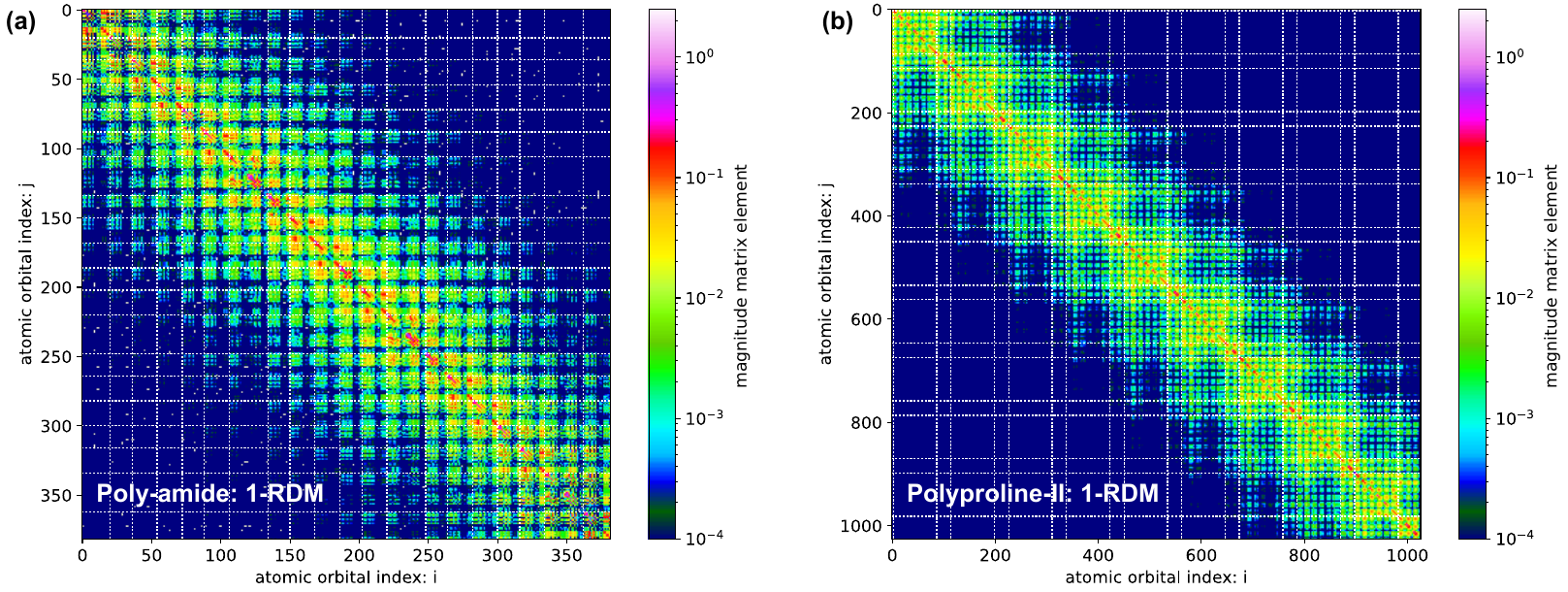}
    \caption{One electron reduced density matrices. Depicted are the 1RDM of poly-amide-3 (panel a) and polyproline-II-9 (panel b) obtained using DFT. Color indicates the magnitude of the matrix elements on a logarithmic scale. White dotted lines indicate the boundaries between building blocks. We note that the building blocks for poly-amide and polyproline-II have different numbers of atomic orbitals and hence the spacing between the white dotted lines is uneven. }
    \label{fig:1RDM_color_PA_PPII}
\end{figure*}

We begin by plotting the 1RDMs for poly-amide and polyproline-II, see Fig.~\ref{fig:1RDM_color_PA_PPII}. These plots complement Fig.~3a of the main text in which we plotted the 1RDMs for polyethylene. Just like for polyethylene, we again observe the rapid decay of matrix elements $\rho^{(1)}_{ij}$ as the building-block distance between $i$ and $j$ increases.

\begin{figure*}[h]
    \centering
    \includegraphics[width=\textwidth]{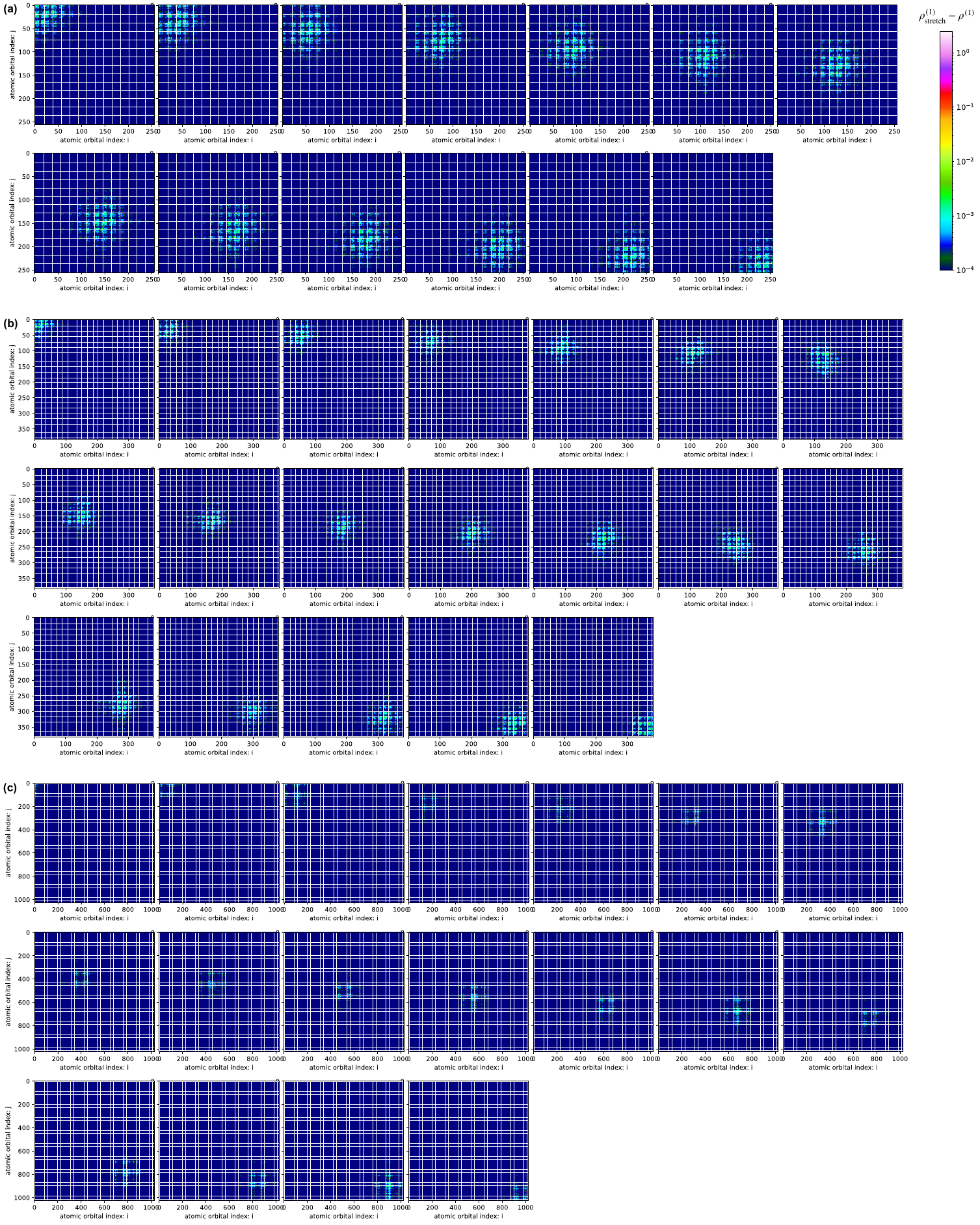}
    \caption{Effect of local perturbations on the 1RDM. Plotted are the differences between the 1RDM of the perturbed molecule, with one of the axial inter-building-block distances stretched by $0.05\,\text{\AA}$, and the 1RDM of the unperturbed molecule. The three series correspond to the three molecules (a) PE-7; (b) PA-3; (c) PP-II-9. In each series the 1$^{text{st}}$ panel corresponds to stretching the distance between building-block 0 and building-block 1, 2$^{\text{nd}}$ panel between building-block 1 and building-block 2, etc.}
    \label{fig:modes}
\end{figure*}

Next, plot a series of figures showing how the 1RDM changes in response to the stretching of one of the axial inter-building-block distances. Each panel of Fig.~\ref{fig:modes} corresponds to the stretching of a different axial distance, with the first panel of Fig.~\ref{fig:modes}a being equivalent to Fig.~3b of the main text. The figure illustrates that 1RDM matrix elements affected by bond stretching are all nearby the bond being stretched. This observation is key to the transferability of our model between different length polymers.

\subsection{Extended plots of predicted vs. computed matrix elements}

\label{sec:som_scatter}

\begin{figure*}[h]
    \centering
    \includegraphics[width=\textwidth]{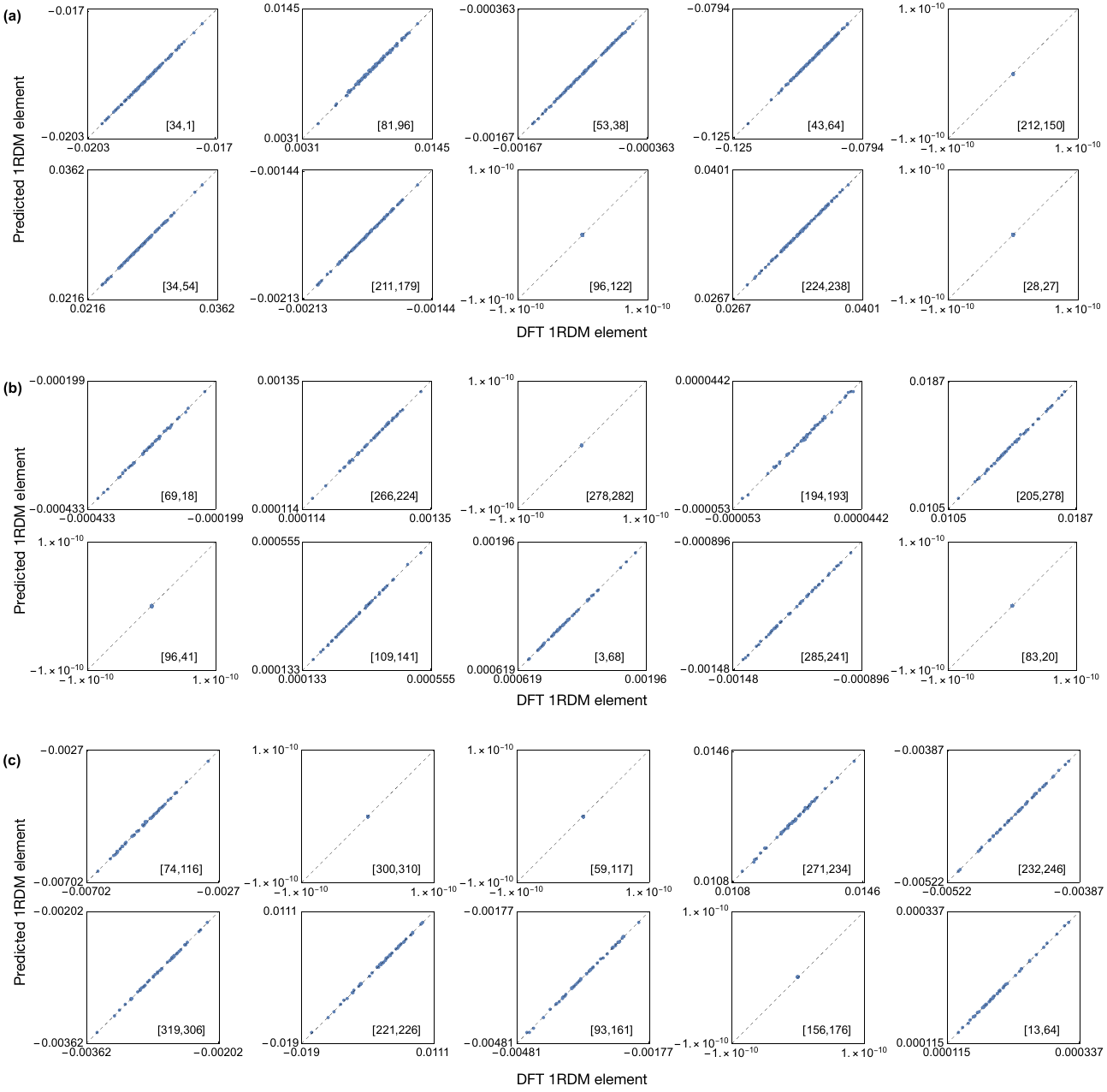}
    \caption{Predicted vs. DFT 1RDM matrix elements for polyethylene: (a) PE-7; (b) PE-8; (c) PE-9. The position of the matrix element in the 1RDM is indicated in each panel in square brackets. }
    \label{fig:PE_scatter}
\end{figure*}

\begin{figure*}[h]
    \centering
    \includegraphics[width=\textwidth]{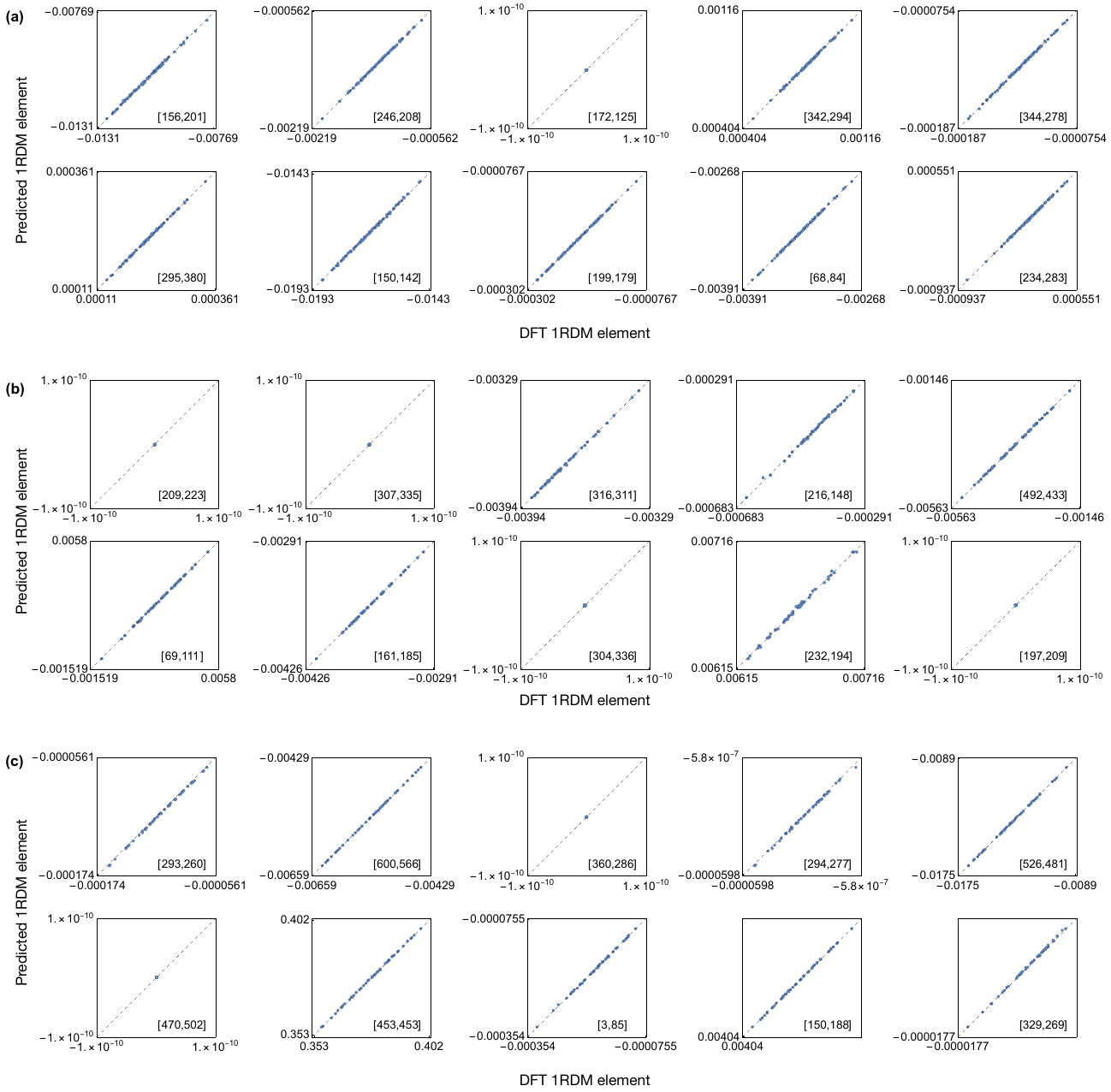}
    \caption{Predicted vs. DFT 1RDM matrix elements for poly-amide: (a) PA-3; (b) PA-4; (c) PA-5. The position of the matrix element in the 1RDM is indicated in each panel in square brackets.}
    \label{fig:PA_scatter}
\end{figure*}

\begin{figure*}[h]
    \centering
    \includegraphics[width=\textwidth]{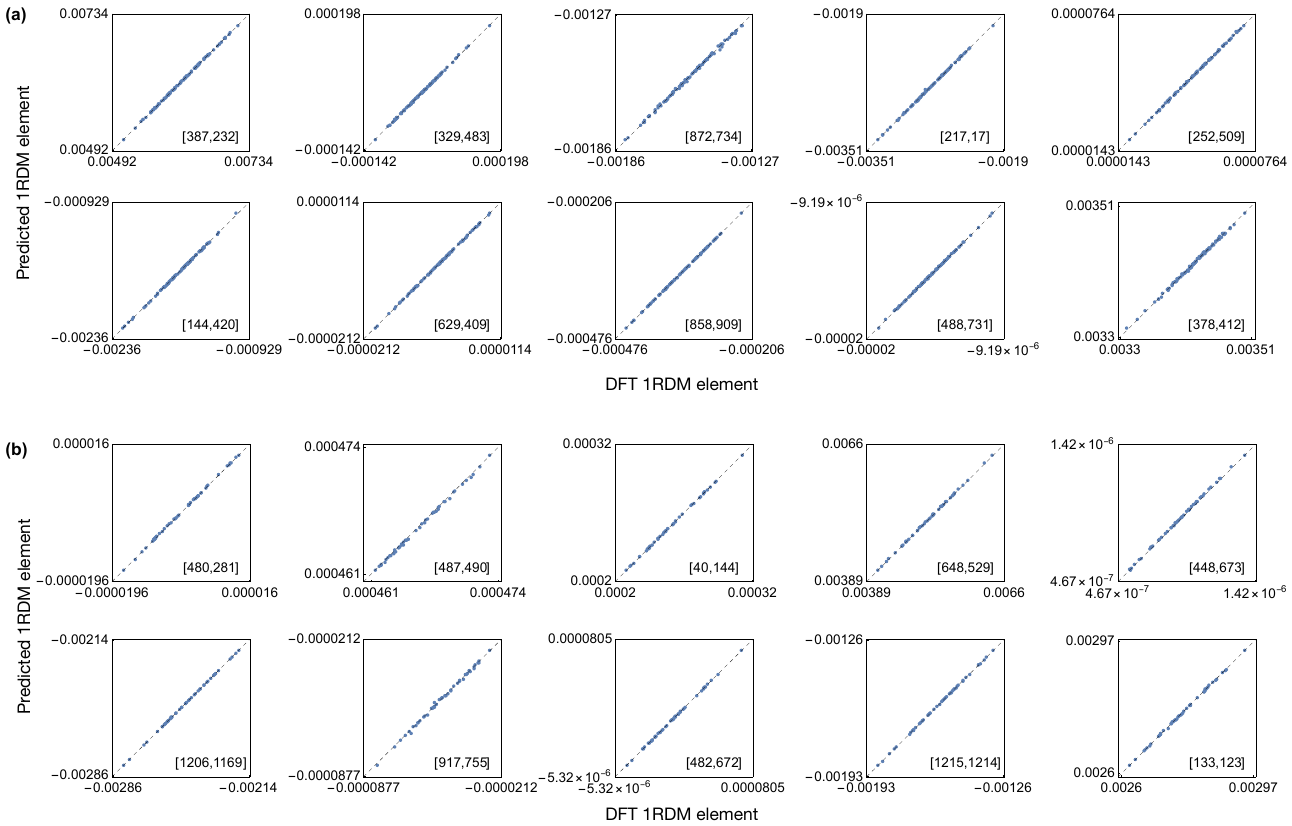}
    \caption{Predicted vs. DFT 1RDM matrix elements for poly-amide: (a) PP-II-9; (b) PP-II-12. The position of the matrix element in the 1RDM is indicated in each panel in square brackets.}
    \label{fig:PPII_scatter}
\end{figure*}

\begin{figure*}[h]
    \centering
    \includegraphics[width=\textwidth]{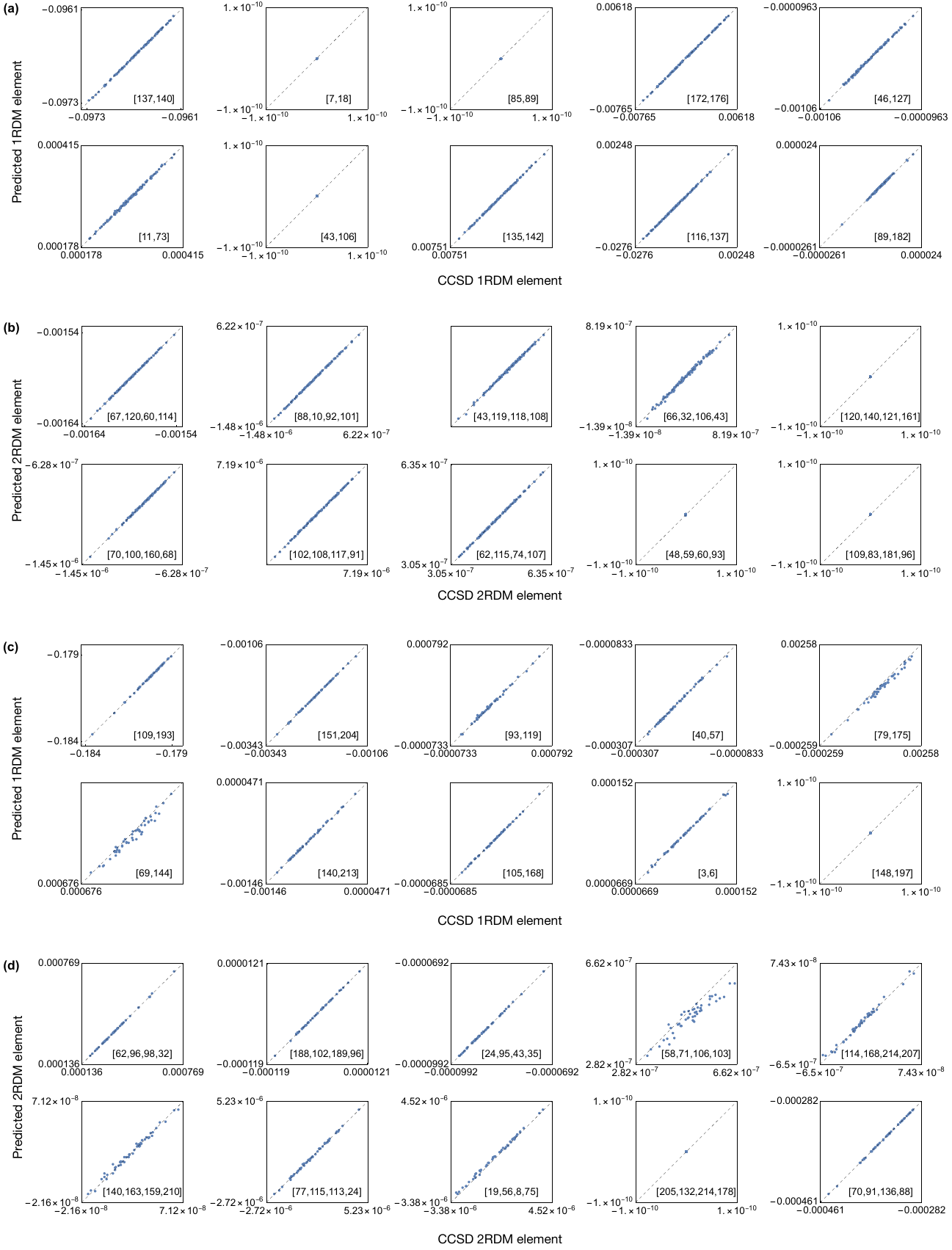}
    \caption{Predicted vs. CCSD matrix elements: (a) PE-5 1RDM; (b) PE-5 2RDM; (c) PE-6 1RDM; (d) PE-7 2RDM. The position of the matrix element in the 1RDM and the 2RDM is indicated in each panel in square brackets.}
    \label{fig:PE_56_scatter_CCSD}
\end{figure*}

In Figs.~\ref{fig:PE_scatter}-\ref{fig:PE_56_scatter_CCSD} we exhibit an extended set of plots of predicted vs computed reduced density matrix elements. Fig.~\ref{fig:PE_scatter} is an extended version of Fig.~5a of the main text and Fig.~\ref{fig:PE_56_scatter_CCSD} is an extended version of Fig.~6a of the main text. Fig.~\ref{fig:PE_scatter} corresponds to polyethylene DFT data, with subfigure (a) for PE-7, subfigure (b) for PE-8, and subfigure (c) for PE-9. The set of panels in each subfigure corresponds to a randomly selected set of matrix elements, with the matrix element labeled in each panel in square brackets. In each panel we plot points for all conformations in the testing set (100 points for PE-7, and 50 points for PE-8 and PE-9).   Fig.~\ref{fig:PA_scatter} and Fig.~\ref{fig:PPII_scatter} are similar to Fig.~\ref{fig:PE_scatter} except they present data for poly-amide and polyproline-II. Fig.~\ref{fig:PE_56_scatter_CCSD} corresponds to polyethylene CCSD data and therefore has subfigures for both 1RDMs and 2RDMs. We note that the scatter in matrix elements of PE-6, depicted in Fig.~\ref{fig:PE_56_scatter_CCSD}c and d, is observably bigger than in matrix elements of other molecules. This observation is consistent with the poorer performance of our model on PE-6 as measured by RMSE of RDM matrix elements and energy error, see Table~1 of the main text. We attribute these shortcomings to the molecules of the training set (PE-5) being too short, see the main text and supplement section ``Model transferability with short polymer training set'' for additional details.

\subsection{Root mean square and normalized root mean square error of predicted matrix elements.}
\begin{figure*} [h]
    \centering
    \includegraphics[width=0.6\textheight]{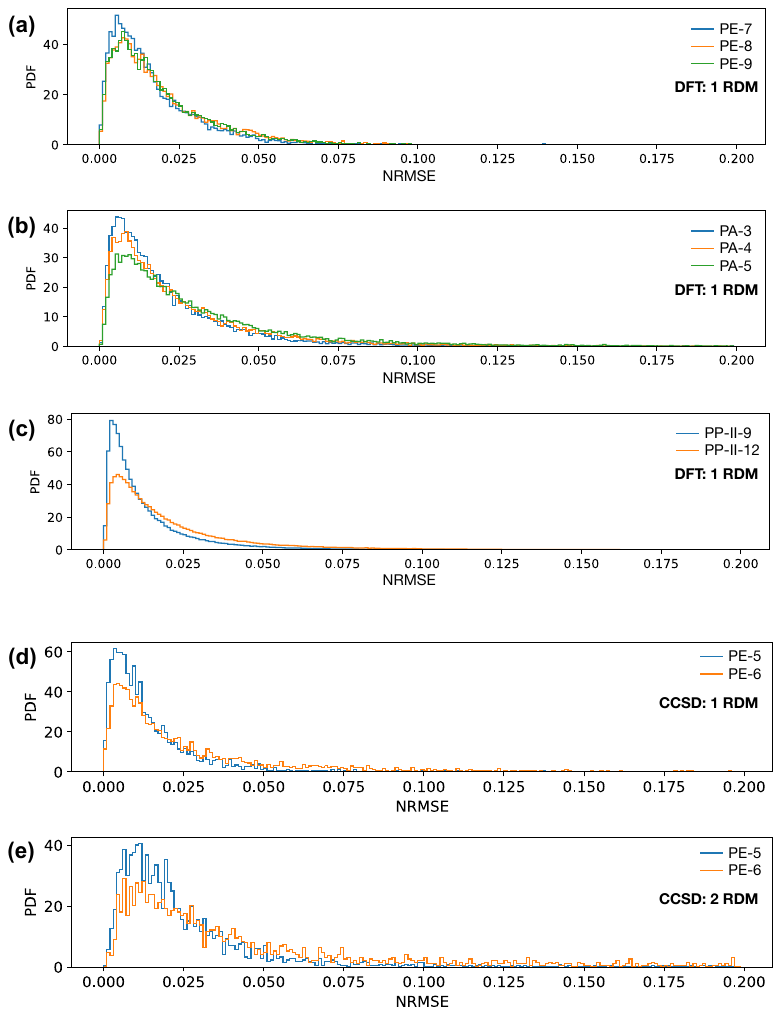}
    \caption{Histograms of the normalized root mean square errors (NRMSE). For each matrix element we computed the NRMSE over all conformations of the molecule in the testing set and plotted the resulting histograms for each testing set. The NRMSE is defined as the RMSE (predicted vs. DFT matrix element) normalized by the standard deviation (DFT/CCSD matrix element). While histograms for the DFT data include all matrix elements that we made predictions for, the histograms for CCSD data are composed of random subsets of 5000 1RDM matrix elements and 5000 2RDM matrix elements. To make normalization work, we dropped matrix elements that were essentially zero as defined by having a standard deviation of less than $10^{-8}$. }
    \label{fig:nrmse_all}
\end{figure*}

\begin{figure*}[h]
    \centering
    \includegraphics[width=0.6\textheight]{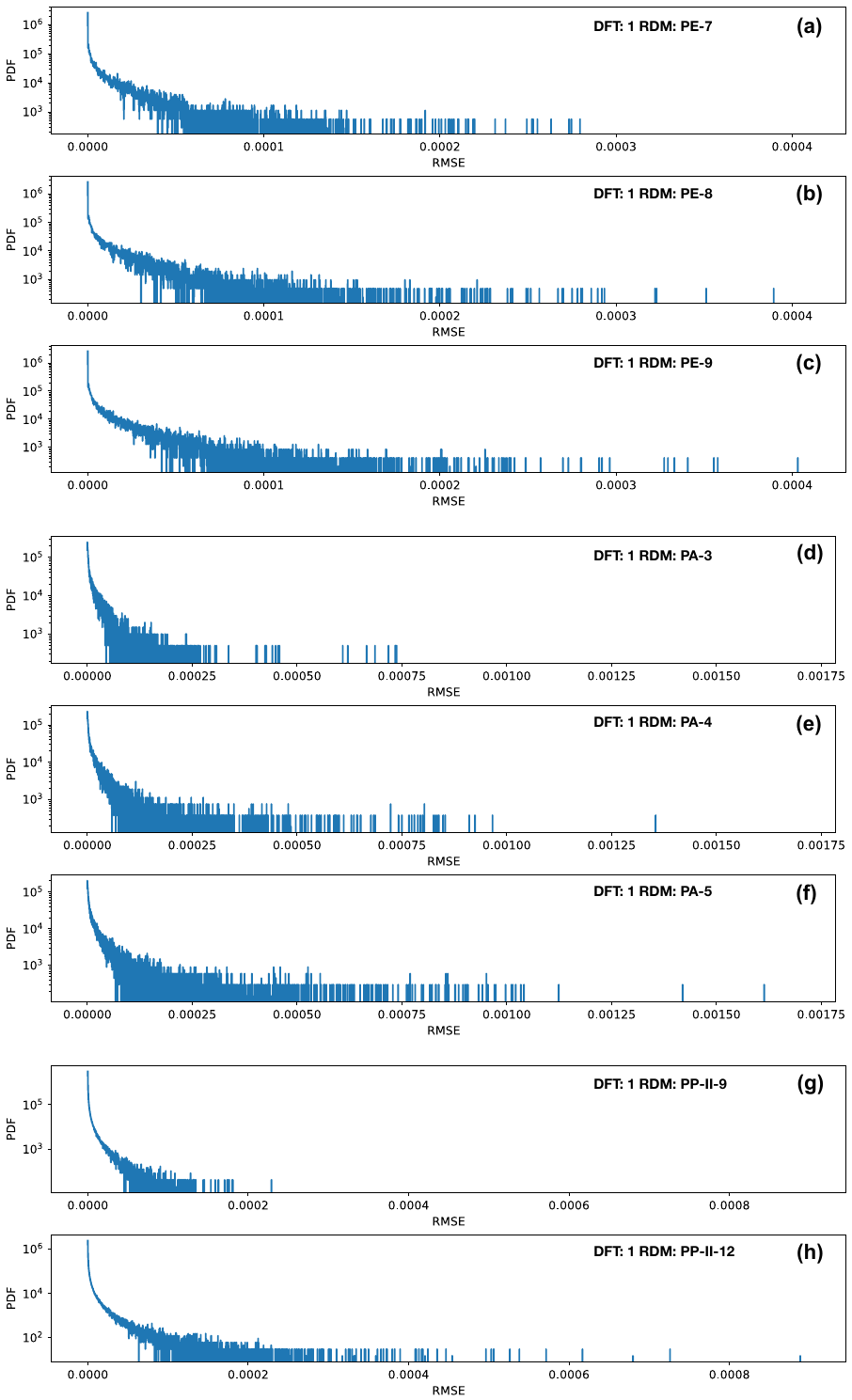}
    \caption{Histograms of the root mean square errors (RMSE) for the DFT datasets. }
    \label{fig:rmse_DFT}
\end{figure*}

\begin{figure*}[h]
    \centering
    \includegraphics[width=0.6\textheight]{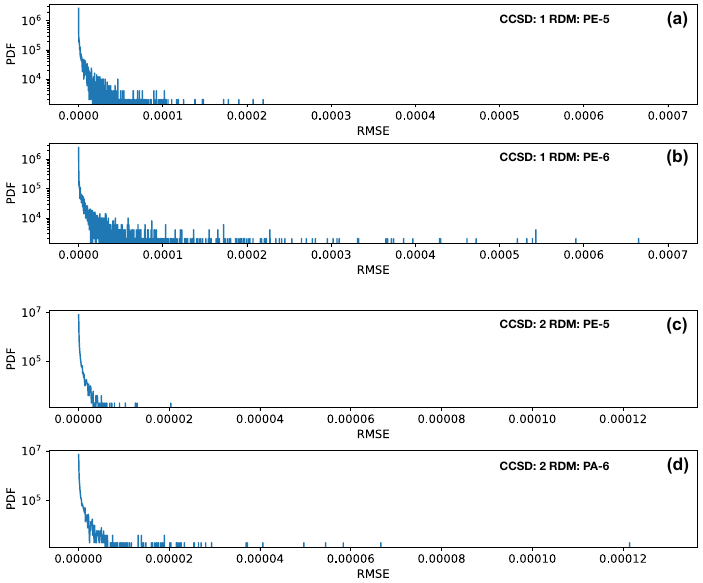}
    \caption{Histograms of the root mean square errors (RMSE) for the CCSD datasets. Histograms were constructed from random subsets of 5000 1RDM matrix elements and 5000 2RDM matrix elements. }
    \label{fig:rmse_CCSD}
\end{figure*}

To quantify the error in the predicted reduced density matrix elements we compute the normalized root mean square error (NRMSE) for the various matrix elements, see Fig.~\ref{fig:nrmse_all}. We observe that in almost all cases the RMSE is a small fraction of the standard deviation, see Table~\ref{table:nrmse}. However, there is a small number of matrix elements for which the RMSE is a significantly bigger fraction of the standard deviation. Fortunately, the un-normalized RMSEs are still quite small, see Fig.~\ref{fig:rmse_DFT} and Fig.~\ref{fig:rmse_CCSD} and cases of large NRMSEs correspond to small standard deviations as opposed to large RMSEs. We do note that NRMSE and RMSE are bigger for PE-6 as compared to the other compounds, see discussion in the previous section, in the main text, and in the supplement section ``Model transferability with short polymer training set''.

\begin{table*}[h]
    \centering
    \begin{tabular}{|c|c|c|c|c|c|c|c|c|c|c|c|}
    \hline
         PE-7 &  PE-8 &  PE-9 &  PA-3 & PA-4 & PA-5 & PP-II-9 & PP-II-12 & PE-5 (1RDM)$^*$ & PE-5 (2RDM)$^*$ & PE-6 (1RDM)$^*$ & PE-6 (2RDM)$^*$ \\
\hline        
         0.017 & 0.02 & 0.02 & 0.21 & 0.026 & 0.031 & 0.015 & 0.025 & 0.015 & 0.026 & 0.033 & 0.077 \\
         \hline
    \end{tabular}
    \caption{Average NRMSE from Fig.~\ref{fig:nrmse_all}. $^*$Averages in the last 4 columns are over 5000 randomly chosen matrix elements as opposed to all predicted matrix elements. }
    \label{table:nrmse}
\end{table*}

\subsection{Model transferability with short polymer training set}
In the main text we ascribe the lower performance of our model in describing PE-6 to a problem of transferability of our model caused by the molecules in training set (i.e. PE-5) being too short. In this section of the supplement we provide additional details supporting these statements.

\begin{figure}
    \centering
    \includegraphics[width=\textwidth]{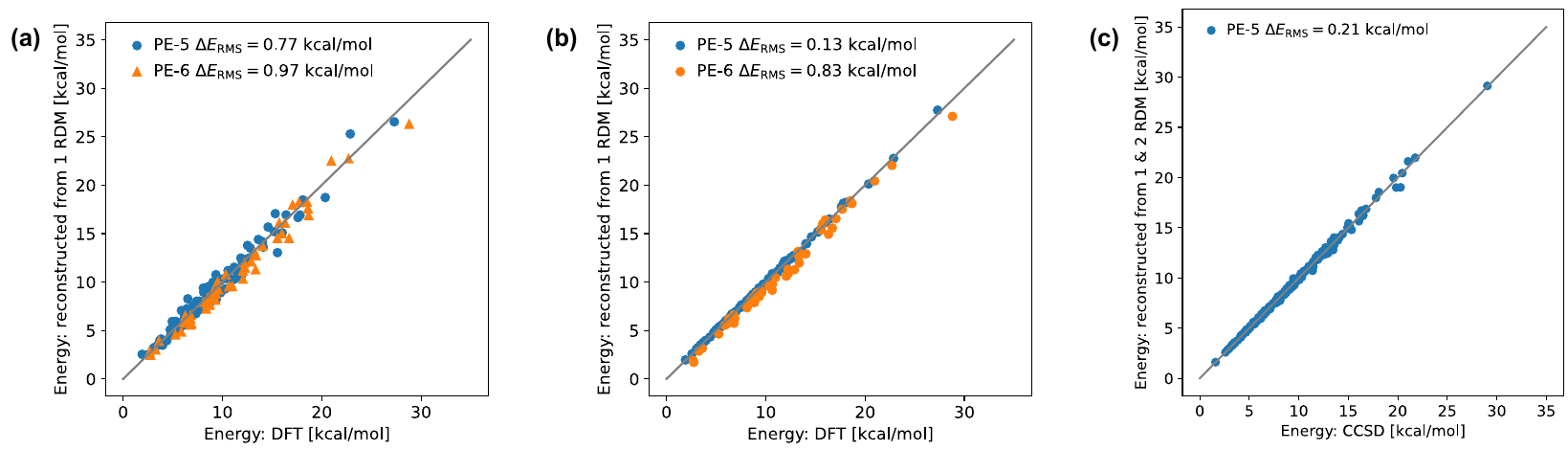}
    \caption{Energy PE-5 and PE-6: machine learned RDMs vs DFT and CCSD obtained RDMs. (a) Machine learning 1-RDMs using short 1-RDM neighborhoods with $r_c=3$ on PE-5 DFT data. (b) Machine learning 1-RDMs using extended 1-RDM neighborhoods on PE-5 DFT data. (c) Machine learning 1- and 2-RDMs with RDM neighborhoods spanning the whole molecule on PE-5 CCSD data.}
    \label{fig:PE_short}
\end{figure}

Since it is computationally challenging to generate CCSD data for longer molecules, we have instead generated an additional DFT data for shorter polymers. Specifically we generated a DFT dataset for PE-5 and PE-6 using the same set of molecular geometries as was used for the CCSD data set. We will now describe how the performance of our models degrades when applying them to the shorter molecules of this new dataset as compared to the longer molecules (PE-7,PE-8, and PE-9) described in the main text.

First, we machine learned a model of the 1RDMs using 1RDM neighborhoods as described in the ``1RDM neighborhoods for DFT data'' part of the ``Methods'' section with $r_q=5$ and $r_c=3$. The value of $r_c$ was limited by the length of PE-5 while the value of $r_q$ was chosen to match CCSD calculations. The performance of this model is depicted in Fig.~\ref{fig:PE_short}a. Compared to a comparable model trained on PE-7 data (see Fig.~5 of the main text), the model trained on PE-5 data performs worse. Performance degradation is observed for molecules of same length as the ones in the training set (PE-5 in the new dataset and PE-7 in the old dataset) as well as in the transferability of the model to longer molecules (PE-6 in the new dataset and PE-8, PE-9 in the old dataset). 

Second, we machine learned a model of the 1RDMs using extended 1RDM neighborhoods as described in the ``1RDM and 2RDM neighborhoods for CCSD data'' part of the ``Methods'' section with $r_q=5$. The performance of this model is depicted in Fig.~\ref{fig:PE_short}b. We see that the use of extended 1RDM neighborhoods significantly improves predictions for PE-5 but only slightly improves predictions for PE-6. This observation indicates that the extended 1RDM neighborhoods significantly improve our model when it is applied to same size molecules as in the training set, but are only marginally helpful in improving the transferability of our model when it is applied to longer molecules. Indeed for same size molecules as in the training set the RMS errors on the energy for PE-5 are smaller than for PE-7 (0.13~kcal/mol vs. 0.22~kcal/mol). However, when we asses transferability we find that RMS errors on the energy of PE-6 are bigger than PE-8 (0.83~kcal/mol vs. 0.32~kcal/mol). Hence, we ascribe the larger errors observed in the main text when we apply our 1- and 2-RDM model learned on PE-5 to PE-6 to a transferability problem caused by the training molecules being too small. 

Finally, we ask how well could our model perform without limitations on the size of RDM neighborhoods? In order to have transferable models it is crucial to have RDM neighborhoods that are edge-like and bulk-like so that a longer molecule can be described by inserting more bulk-like neighborhoods between the edge-like neighborhoods. However, if we drop the requirement for transferability, our neighborhoods can span the entire molecule. We have machine learned the CCSD 1RDM and 2RDM data for PE-5 and plotted the results in Fig.~\ref{fig:PE_short}c (since we dropped transferability the figure has no data for PE-6). Although this is the same CCSD dataset as we used in the main text, we used 800 geometries for training and 200 for testing. The resulting RMS error on energy is only 0.21~kcal/mol. This level of error indicates the expected performance of the machine learning model when it is not limited by the RDM neighborhood size.

\end{document}